\documentclass[10pt,twocolumn]{article}
\usepackage{epsfig}
\usepackage{subfigure}
\usepackage{algorithm}
\usepackage{algorithmic}
\usepackage{times}
\usepackage{url}

\newcommand{\myitem}[1]{\noindent\vspace*{1mm}\hspace*{7mm}\parbox{3in}{$\bullet$~#1}\vspace*{1mm}\\}
\begin{document}

\title{ {\Large\bf Protecting Public-Access Sites \\
Against Distributed Denial-of-Service Attacks} }

\author{Katerina J. Argyraki \hspace{1cm} David R. Cheriton\\
Distributed Systems Group\\
Stanford University\\
\{argyraki, cheriton\}@dsg.stanford.edu}

\date{}
\maketitle

\begin{abstract}
A distributed denial-of-service (DDoS) attack can flood 
a victim site with malicious traffic, 
causing service disruption or even complete failure. 
Public-access sites like amazon or ebay are particularly vulnerable 
to such attacks, because they have no way of a priori blocking
unauthorized traffic.

We present \emph{Active Internet Traffic Filtering} (AITF),
a mechanism that protects public-access sites from highly distributed attacks
by causing undesired traffic to be blocked as close as possible to its sources.
We identify filters as a scarce resource and 
show that AITF protects a significant amount of the victim's bandwidth, 
while requiring from each participating router a number of filters 
that can be accommodated by today's routers.
AITF is incrementally deployable, because 
it offers a substantial benefit even to the first sites that deploy it.
\end{abstract}

\section{Introduction}
\label{intro}

Distributed denial-of-service (DDoS) attacks are recognized as one of the 
most dangerous threats to Internet operation.
An attacker typically uses self-propagating code to compromise a large number 
of vulnerable hosts and commands them to flood the victim with malicious 
traffic.
Any organization or enterprise that is dependent on the Internet
can be subjected to such an attack,
causing its service to be severely disrupted, if not fail completely.

Current responses to DDoS attacks usually involve manual 
``hop-by-hop'' filter propagation 
that stops at arbitrary points in the network:
Typically, the operator of a site under attack identifies the nature 
of attack traffic and installs filters in the site's
firewall to block this traffic.
Then she contacts her ISP and has the ISP install comparable filters in its
routers in order to protect the tail circuit to the site.
The ISP may contact the next provider upstream and so on.

The first two problems with the current approach 
have been identified and addressed in the literature:

First, the current approach is manual and, hence, 
too slow to keep up with a sophisticated attack.
Mahajan {\it et al.} have proposed \emph{Pushback}, 
a mechanism that automates filter propagation:
A site under attack sends a request to its edge-router
to block (or rate-control) attack traffic.
The edge-router satisfies the request and may as well
identify the upstream routers that contribute most
attack traffic and propagate similar requests to them~\cite{Mahajan01}.

Second, the current approach is ineffective in the face
of source address spoofing.
I.e., an attack source can spoof multiple IP addresses and 
use them to attack the victim site.
In that case, there is practically no way for the victim site 
to distinguish attack traffic from ``good'' traffic.
Thus, blocking (or rate-controlling) attack traffic by source IP address
leads to sacrificing most good traffic as well.
To address this, researchers have proposed anti-spoofing mechanisms
that provide each packet with a non-spoofable identifier~\cite{Hamadeh03,Yaar03}.
These identifiers can be used by the victim site to identify
and selectively block attack traffic.

In this paper, we identify and address a new challenge in protecting 
public-access sites against DDoS attacks:
Blocking attack traffic using a number of filters per router
that can be accommodated in today's routers,
i.e., efficiently managing the ``filtering capacity'' 
of the Internet during a DDoS attack.

Each hardware router has only a limited number of filters
that can block traffic without degrading the router's performance 
(wire-speed filters).
The limitation comes from cost and space:
Wire-speed filters are typically stored in TCAM 
(Ternary Content Addressable Memory),
which they share with the router's forwarding table.
Some of the largest TCAM chips available today accommodate $256$K entries
\cite{ultra18} and cost about \$$200$~\cite{elecdesign}.
A sophisticated router linecard fits at most $1$ TCAM chip
\cite{cisco} i.e., 
at most $256$K filters per network interface.

On the other hand,
self-propagating code can compromise $1$ million hosts \cite{Stanisford02}
and synchronize them to attack a certain victim.
The victim may use packet marking to identify $1$M attack flows,
and automatic filter propagation to send out $1$M filtering requests 
to block these flows.
However, all these requests must be propagated through the 
victim's edge-router and the Internet core.
$1$M filters is too high a load for any edge-router.
Moreover, $1$M filters \emph{per potential victim} is too 
high a load for the Internet core.

Yet, there \emph{are} enough filtering resources in the Internet
to block such large-scale attacks.
For example, an attack coming from $100$K different networks involves 
at least $100$K routers, i.e., about $25$ billion filters are available to 
help block attack traffic.
Clearly, the closer we get to the attack sources, 
the larger the amount of filtering resources available 
per attack source;
it is the victim's edge-router and the Internet core that are the
``filtering bottleneck".
Therefore,
our solution focuses on causing attack traffic to be blocked
as close as possible to its sources,
while requiring a reasonable number of filters from each
participating (edge or core) router.

In this paper, we present AITF (Active Internet Traffic Filtering),
a mechanism that protects public-access sites against highly 
distributed denial-of-service attacks.
AITF requires a bounded amount of resources from each participating router,
on the order of a few thousand entries of TCAM memory and a few gigabytes of DRAM memory --
we choose these requirements based on real products~\cite{cypress, cisco}.
AITF is incrementally deployable,
because it offers substantial benefit even to
the first sites that deploy it.

AITF assumes that there exists some anti-spoofing mechanism, 
which limits source address spoofing \emph{to a certain degree}.
The accuracy of the anti-spoofing mechanism affects 
AITF's power.
However, it does not affect AITF's resistance to malicious abuse.
I.e., AITF prevents, with high probability, a malicious node from
forging filtering requests that would interrupt other nodes' communications;
this probability is independent from the underlying anti-spoofing mechanism.

The paper is organized as follows:
Section \ref{assumptions} describes our assumptions and
section \ref{protocol} describes the protocol in detail. 
Section \ref{estimates} presents performance estimates
and section \ref{deployment} discusses Internet deployment.
Section \ref{sim} presents simulation results.
Sections \ref{issues} and \ref{related}
discuss vulnerability issues and related work, 
and section \ref{conclusion} concludes.

\section{Assumptions}
\label{assumptions}

\subsection{Limited Source Address Spoofing}

We do not assume that source address spoofing is $100$\% eliminated;
we assume the existence of some anti-spoofing mechanism,
which limits source address spoofing to a certain degree.
However, we define a minimal anti-spoofing requirement,
which is necessary for AITF to operate: 
\emph{Any end-host or router can inspect a packet and identify, at wire-speed, 
and with high probability, 
a border router that forwarded the packet and is located close to its source.}

Clearly, the current Internet architecture does not meet our minimal requirement.
However, a lot of research effort has been invested in designing
deployable mechanisms that do:
Park {\it et al.} have proposed \emph{Distributed Packet Filtering} (DPF),
where each participating router inspects the traffic that arrives
at each interface and drops all packets with detectably fake 
IP source addresses~\cite{Park01}.
Hamadeh {\it et al.} have proposed a packet marking scheme,
where each packet is ``stamped'' by the first participating 
border router that forwards the packet~\cite{Hamadeh03}.
Yaar {\it et al.} have proposed \emph{Path Identifier} (Pi),
where each participating router marks the forwarded packets,
so that each packet obtains a ``fingerprint'' that reflects
the entire path followed by the packet~\cite{Yaar03}.
Each of these mechanisms either meets our minimal requirement
or can be slightly modified to do so.

As we will see,
this minimal requirement is enough for AITF to mitigate 
large DDoS attacks.
However, AITF's power increases,
if the underlying anti-spoofing mechanism provides
(i) higher accuracy in identifying the sources of attack traffic, and
(ii) more complete information about the path followed by attack traffic.

\subsection{Detection and Identification of Attack Traffic}

We assume that a public-access site can detect when it is under attack.
We also assume that, using an anti-spoofing mechanism,
the site can identify the set of traffic flows that
contribute most to the attack and describe each one of them with
an appropriate flow label.
The format of the flow label depends on the anti-spoofing mechanism.
For example, for DPF, it is simply an \{IP destination prefix, IP source prefix\} pair;
for Pi, it is an \{IP destination prefix, $16$-bit path identifier\} pair.

If an edge-network that hosts an attack source prevents its clients from spoofing,
the traffic sent by the attack source can be specified and blocked
by the victim as an individual flow.
However, if the attack source is allowed to spoof multiple addresses,
its traffic can only be specified and blocked by the victim as an aggregate
of traffic forwarded by the same border router.
Therefore, an edge-network that hosts attack sources and allows local 
spoofing will suffer coarser-grained filtering of its traffic.

\section{Protocol Description}
\label{protocol}

\subsection{Overview}
\label{overview}

Upon determining that it is under attack,
the victim sends a filtering request to its gateway.
The victim's gateway temporarily blocks attack traffic and
identifies a border router located close to the attacker -- 
call it the \emph{attacker's gateway}.
The victim's gateway initiates a ``counter-connection setup''
with the attacker's gateway, i.e.,
an agreement not to transmit certain packets --
the same way a TCP connection setup is an agreement to
exchange packets.
If the attacker's gateway does not respond or does not respect the agreement, 
the mechanism \emph{escalates} to the second round.

In the second round, the victim's gateway blocks all traffic
forwarded by the attacker's gateway to the victim.
It does that (i) either by blocking the traffic locally, or
(ii) by contacting another border router close to the attacker.
Escalation can continue until a router along the attack 
path responds and a ``counter-connection setup'' is completed.
If no router responds, attack traffic is blocked locally by the victim's gateway.
However, as we will see, AITF both assists and motivates
routers close to the attacker to help block attack traffic.

Note that our minimal requirement allows only for one escalation round,
because the victim's gateway can identify only one 
border router located close to the attacker.

\subsection{AITF Messages and Parameters}
\label{messages}

An AITF message includes the following fields:

\begin{itemize}
\item{The \emph{flow labels} field specifies a set of undesired traffic flows
that the sender wants blocked.}
\item{The \emph{SYN} field and the \emph{ACK} field are $1$-bit flags.}
\item{The \emph{nonce} field is a random number.}
\end{itemize}

Borrowing from TCP terminology, 
we use the terms ``SYN message'', ``ACK message'' and ``SYN/ACK message''
to refer to an AITF message with just the SYN flag set, 
just the ACK flag set, or both flags set, respectively.

There are $2$ AITF parameters:

\begin{itemize}
\item{The \emph{temporary filter timeout} $T_{tmp}$.}
\item{The \emph{filtering window} $T$.}
\end{itemize}

\subsection{Incremental Algorithm Description}
\label{single}

\subsubsection{Terminology}

A \emph{filtering request} is a request to block an undesired flow
described by a specific flow label.
An AITF message may include more than one filtering requests.
An \emph{AITF router} is a router that runs the AITF algorithm.

We define the following terms with respect to an undesired traffic flow:

\begin{itemize}
\item{The \emph{attacker} is the source of the undesired flow.}
\item{The \emph{victim} is the detonation of the undesired flow.}
\item{The \emph{attack path} consists of the attacker, 
the set of AITF routers the undesired flow goes through, and the victim.}
\item{The \emph{attacker's gateway} is the AITF router closest to the attacker
along the attack path.}
\item{The \emph{victim's gateway} is the AITF router closest to the victim
along the attack path.}
\end{itemize}

\subsubsection{Basic Algorithm}
\label{basic}

\begin{enumerate}

\item{The victim sends a filtering request to the victim's gateway,
specifying an undesired flow $F$.}

\item{The victim's gateway ($vGw$):
\begin{enumerate}
\item{If the victim has exceeded a configured maximum 
filtering request rate $R$,
the victim's gateway drops the message. Else:}
\item{Installs a temporary filter to block $F$ for $T_{tmp}$ time units
and updates the victim's filtering request rate.}
\item{Initiates and completes a 3-way handshake with the attacker's gateway.}
\end{enumerate}
}

\item{The attacker's gateway ($aGw$):
\begin{enumerate}
\item{Responds to the 3-way handshake.}
\item{Upon completion of the 3-way handshake,
installs a filter to block $F$ for $T_{tmp}$ time units.}
\item{Sends a filtering request to the attacker,
specifying $F$ as the undesired flow.}
\end{enumerate}
}

\item{The attacker stops $F$ for $T \gg T_{tmp}$ time units 
or risks being disconnected by its gateway $aGw$.}

\end{enumerate}

The victim's gateway installs a filter only temporarily,
in order to immediately protect the victim until the attacker's
gateway takes responsibility.
Similarly, the attacker's gateway installs a filter temporarily,
in order to immediately block the undesired flow until the
attacker stops.
As we will see,
spending a filter for $T_{tmp}$ time units in order to block an
undesired flow for $T \gg T_{tmp}$ units is a key feature that
enables AITF to block large attacks with a reasonable number 
of filters per router.

\subsubsection{The 3-way Handshake}
\label{handshake}

The handshake involves the following messages:
\begin{enumerate}
\item{The victim's gateway sends a SYN message to the attacker's gateway.}
\item{The attacker's gateway responds with a SYN/ACK message addressed \emph{to the victim}.}
\item{The victim's gateway intercepts the SYN/ACK and responds with an ACK.}
\end{enumerate}

The handshake prevents a malicious router $M$ from causing 
a filter to be installed at router $aGw$ and block traffic 
to $V$ without $V$'s cooperation:
When $aGw$ receives an AITF SYN that requests to block traffic to $V$,
$aGw$ responds with a SYN/ACK that includes a nonce and is addressed to $V$.
Malicious router $M$ cannot snoop that SYN/ACK, 
unless it is located on the path between $aGw$ and $V$.
$M$ must send an ACK with the correct nonce value, otherwise its request is rejected.
By picking a sufficiently large and properly random value for the nonce, 
it can be made arbitrarily difficult for $M$ to guess the nonce.  

The 3-way handshake does open the door to certain abuses.
We discuss them in section \ref{handabuse}.

\subsubsection{Shadow Filtering Table}
\label{shadow}

In the basic algorithm,
the attacker's gateway installs a filter 
and blocks the undesired flow $F$ for $T_{tmp}$ time units,
but expects the attacker to stop sending $F$ for $T \gg T_{tmp}$ time units.
A smart attacker can play an ``on-off'' game:
pretend to cooperate, pause $F$ and resume it as soon 
as the attacker's gateway removes its temporary filter.

To address this case, the attacker's gateway
records a shadow of the temporary filter in DRAM for $T$ time units.
If the attacker's gateway receives a second filtering request for $F$
before the shadow filter expires, 
the attacker's gateway installs a temporary filter as before.
This time, however, if the filter catches $F$ traffic,
the attacker's gateway disconnects the attacker.

The same technique is applied by the victim's gateway 
to verify that the attacker's gateway respects their
filtering agreement, i.e., keeps the undesired flow blocked.

Note that an AITF router must ``remember'' a flow
for as long as the flow must be filtered.
That is, AITF does not reduce the amount of state required for filtering,
it simply ``moves'' it from TCAM to DRAM.
This makes filtering cheaper,
given that DRAM costs about \$$0.5$/MB, 
while TCAM costs about \$$200$/MB.

\subsubsection{Escalation}
\label{escalation}

In the basic algorithm, an undesired flow is blocked
only when the attacker's gateway cooperates.
It is possible that the attacker's gateway either does not
respond at all to the 3-way handshake or
responds and then fails to keep the undesired flow blocked.

If the attacker's gateway does not respond to the 3-way
handshake within a given grace period,
the mechanism escalates to the next round.

If the attacker's gateway does respond,
but fails to block the undesired flow for $T$ time units:
\begin{enumerate}
\item{The victim's gateway re-initiates the 3-way handshake.} 
\item{If the undesired flow still does not stop, 
the mechanism escalates to the next round.}
\end{enumerate}
Step $1$ gives a second chance to the attacker's gateway to block
the undesired flow, in case the attacker is playing the on-off
game described in section \ref{shadow}.

\begin{figure}[htbp]
\centering
\includegraphics[width=9cm, height=4cm]{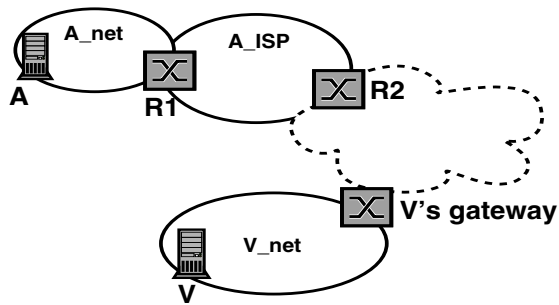}
\caption{\small Escalation Example - 
In round $1$, the victim's gateway asks from router $R_1$ to block
all traffic from attacker $A$ to victim $V$.
In round $2$, the victim's gateway asks from router $R_2$ to block
all traffic from router $R_1$ to the victim $V$.}
\label{escfig}
\end{figure}

When the mechanism escalates,
the attacker's gateway is viewed as an attacker
i.e., all traffic forwarded by the attacker's gateway to the victim is
now considered an attack flow that must be blocked.
So, every time the mechanism escalates, filtering becomes more aggressive.
Escalation involves two potential actions:
\begin{enumerate}
\item{The victim's gateway blocks all traffic from the attacker's gateway
to the victim locally. Or:}
\item{The victim's gateway initiates a 3-way handshake with the
AITF located closest to the attacker's gateway on the attack path.}
\end{enumerate}
If the victim's gateway has the necessary filtering resources to filter 
the attacker's gateway locally, it chooses action $1$.
Otherwise, it chooses action $2$,
which results in the basic algorithm being replayed,
with the attacker's gateway taking the role of the attacker.
This is illustrated in figure \ref{escfig}.

We repeat that our minimal anti-spoofing requirement
enables only choice $1$,
because contacting a second border router close to
the attacker requires that the victim (or its gateway) have
more information on the attack path.

\subsection{AITF Parameter Values}
\label{parameters}

\subsubsection{Temporary Filter Timeout}

The goal of the temporary filter on the victim's gateway is to
block attack traffic until the 3-way handshake is complete.
Considering that Internet round-trip times 
range from $50$ to $200$ msec,
a safe value for $T_{tmp}$ is $1$ sec.

\subsubsection{Filtering Window}

A positive aspect of AITF is that the administrator of an attack source
immediately learns that their machine has been compromised 
and is being used to send undesired traffic --
she finds out either by reading the filtering requests sent to the machine
or by noticing that the machine has been disconnected.
So, the choice of the filtering window $T$ involves the following trade-off:
A large $T$ of, say, $30$ minutes guarantees that, 
once a filtering request has been satisfied, the victim will not
receive any traffic from the corresponding attack source for at least $30$ minutes. 
However, it also guarantees that the attack source will not
send any traffic at all for $30$ minutes, 
even if it gets immediately patched and cleaned by its administrator.

For the rest of the paper, we use example values of $T=2$ or $10$ min --
although our performance estimates are always presented as functions of $T$.
We realize that it is debatable whether $2$ or even $10$ minutes are enough
for an administrator to read a filtering request and patch their system.
We use these example values only to illustrate our performance estimates
with specific numbers, and leave the optimal $T$ value as an open question. 

\section{Performance Estimates}
\label{estimates}

We use the following metrics to evaluate AITF:

\begin{itemize}

\item{ 
The {\bf filtering response time} with respect to an undesired flow $F$ 
is the amount of time that elapses from the moment the victim sends 
the first AITF message with flow label $F$ until $F$ is blocked.
}

\item{
The {\bf lost victim bandwidth} is the fraction of the
victim's bandwidth that gets consumed by undesired flows.
The {\bf preserved victim bandwidth} is the fraction of the
victim's bandwidth that remains available to legitimate flows.
}

\item{
The {\bf filtering resources} include 
the number of filters (the amount of TCAM memory) 
and the number of shadow entries (the amount of DRAM memory)
required by the AITF algorithm.
}

\end{itemize}

\subsection{Victim's per Flow Perspective: Filtering Response Time}
\label{frt}

In the common case,
filtering response time for undesired flow $F$
equals the one-way delay $D$ from the victim to its gateway,
which is typically a few milliseconds.

Spikes can still occur after $D$,
if the attacker or the attacker's gateway play the on-off game described
in section \ref{shadow}.
A router that plays this game would necessarily have to be controlled by the attacker.
Each such compromised router along the attack path induces two spikes, 
spaced out by $T_{tmp}$ time units,
because each router is given two chances to have the undesired flow blocked.
The effect of these spikes on the victim is insignificant,
as we demonstrate in our simulation results.

\subsection{Victim's Aggregate Perspective: 
Preserved Victim Bandwidth}
\label{eab}

One filtering request causes an undesired flow to be blocked 
for $T$ time units, where $T$ is the filtering window.
Thus, a victim allowed to send up to $R$ filtering requests per time unit
can have $R\cdot T$ simultaneous undesired flows blocked.

It follows that, the lost victim bandwidth
is equal to the bandwidth $B_{\mathit {att}}$ of all the
undesired flows $N_{att}$ minus the bandwidth of the $R\cdot T$
undesired flows that get blocked.
The preserved victim bandwidth $B_{\mathit {p}}$
is equal to the total victim bandwidth $B_{\mathit {v}}$
minus the lost victim bandwidth:

\[ B_{\mathit{p}} \ge \left\{ \begin{array} {l l}
B_v - B_{\mathit{att}} (1-\frac{R\cdot T}{N_{\mathit{att}}}) & 
{\mathit {if}} R\cdot T < N_{\mathit{att}} \\
B_v & {\mathit {if}} R\cdot T \ge N_{\mathit{att}} 
\end{array} \right. \]

For example, 
consider a victim site that connects to the Internet through a $100$ Mbps link,
and a $100$ Mbps attack coming from $1$M attack sources.
Without AITF, this attack would completely consume the victim's bandwidth.
For $R = 1$K filtering requests per second and $T = 10$ minutes, 
the victim preserves $60$ Mbps of its bandwidth.
For $R = 2$K filtering requests per second and the same $T$,
the victim preserves almost $100$\% of its bandwidth.

A proof for the above formula can be found in the appendix.

\subsection{Router's Perspective: Filtering Resources}
\label{fr}

Consider AITF router $Gw$, 
which connects edge-network $Net$ to its ISP provider.
$Gw$ needs filters to run both the attacker's gateway algorithm -- 
i.e., block undesired traffic sent by $Net$ hosts --
and the victim's gateway algorithm -- 
i.e., block undesired traffic sent to $Net$ hosts.

The attacker's gateway algorithm spends a filter on each attacker 
only temporarily.
Beyond that, either the attacker stops or has its port disabled.
Thus, $Gw$ needs a number of filters proportional to the 
number of connected hosts.

The victim's gateway algorithm initially spends one filter per filtering request
for only $T_{\mathit {tmp}}$ time units.
Therefore, in order to satisfy $R_{\mathit {max}}$ filtering requests per time unit,
$Gw$ needs at least $R_{\mathit {max}}\cdot T_{\mathit {tmp}}$ filters.
In the best-case scenario, all of the attacker gateways cooperate,
and $Gw$ does not spend any more filters.
In the worst-case scenario, none of the attacker gateways respond,
and $Gw$ ends up filtering traffic from all of them.
The total number of filters required by the victim's gateway algorithm is
bounded by these two cases:
$$R_{\mathit {max}}\cdot T_{\mathit {tmp}} \le N_{\mathit {filters}} \le N_{\mathit {agw}}$$
where $R_{\mathit {max}}$ is the max rate of satisfiable filtering requests,
and $N_{\mathit {agw}}$ is the number of attacker gateways.

For example, 
consider that one of $Gw$'s clients is receiving $1$M undesired flows 
forwarded by $N_{\mathit {agw}} = 160$K\footnote{
$160$K is the total number of network prefixes currently advertised in the Internet
\cite{sprintlabs},
so this is an upper bound for the number of attacker gateways.}
attacker gateways.
To preserve all its bandwidth,
the victim needs to send $R = 2$K filtering requests/sec (see section \ref{eab}).
Thus, if all attacker gateways cooperate, 
$Gw$ must use $2$K filters to preserve $100$\% of its client's bandwidth.
If nobody cooperates, $Gw$ must use $160$K filters to have the same effect.

Apart from wire-speed filters, the AITF algorithm also requires
DRAM for the shadow filtering table.
Each received filtering request is stored in DRAM for $T$ time units.
Thus, a router needs $R_{\mathit {max}}\cdot T$ DRAM entries,
where $R_{\mathit {max}}$ is the maximum rate of filtering requests
that can be satisfied.
Note that DRAM is not the limiting factor:
Even if a network were attacked by $10$M attack sources at the same time,
a few gigabytes of DRAM would be enough for the shadow table.

\section{Internet Deployment}
\label{deployment}

\subsection{AITF Domains}
\label{domains}

Rather than requiring every Internet router to support AITF, 
it is sufficient for the border routers between administrative domains to support it.  
We introduce the notion of an \emph{AITF domain} 
as an administrative domain whose border routers 
are AITF routers.

An AITF domain has a \emph{filtering contract} with each local end-host
and peering domain.
Such a contract specifies a maximum \emph{filtering request rate}, i.e.,
the maximum rate at which the AITF domain can send/receive
requests to block undesired flows to/from each end-host and peering domain.
An AITF domain enforces the specified rates and
indiscriminately drops messages from an end-host/domain
when that end-host/domain exceeds the agreed rate.

In a way,
an AITF domain is the ``dual'' of a BGP Autonomous System (AS):
ASs exchange routing information,
which communicates their willingness to relay certain packets.
Similarly,
AITF domains exchange filtering information,
which communicates their unwillingness to receive certain packets.
However, an AITF domain differs from an AS, in that it exchanges
messages with other AITF domains that are not adjacent to it --
recall that the victim's gateway talks directly to the attacker's gateway.
It seems natural for every Autonomous System (i.e., every ISP, national network
and international backbone), to map to a separate AITF domain.

Our position is that the filtering contract should be part of the
Service Level Agreement (SLA) signed between the customer
and the service provider.
In this way,
when a domain agrees to provide a certain amount of bandwidth to a customer,
the provider also agrees to satisfy a certain rate of filtering requests
coming from that customer.
At the same time, the customer agrees to satisfy a certain
rate of filtering requests coming from the provider.
The customer-provider pair can be an end-host and an edge-network,
or an edge-network and an ISP,
or even a small ISP and a backbone network.

\subsection{Initial Deployment}
\label{initial}

AITF effectiveness increases with the number of administrative domains that deploy it.
For example, in section \ref{fr} we saw that the victim's gateway can block
$1$M attack flows using only $2$K filters, 
provided that all the attacker gateways cooperate.
However, in order for AITF to become widely deployed in the first place,
it has to offer a substantial benefit even to the first domains that deploy it.
In this section, we illustrate that this is the case using an example.

\begin{figure}[htbp]
\centering
\includegraphics[width=9cm, height=4cm]{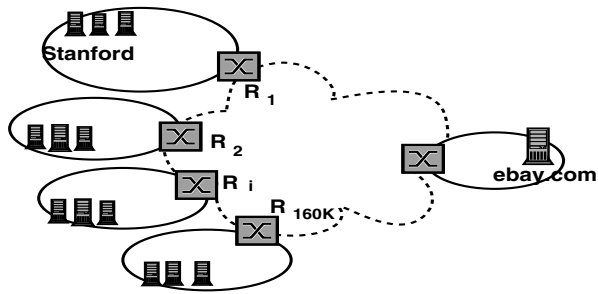}
\caption{\small Initial Deployment - 
only the ebay.com site and Stanford University have deployed AITF.}
\label{initfig}
\end{figure}

In figure \ref{initfig}, only two networks in the entire Internet have deployed AITF:
Popular website ebay.com and Stanford University.
A worm compromises $1$M hosts, distributed among $160$K edge-networks,
and commands them to launch a DDoS attack against ebay.com (66.135.192.87).
$160$K is the total number of prefixes currently advertised in the Internet,
and, thus, an upper bound for the number of edge-networks 
hosting attack sources~\cite{sprintlabs}.
Therefore, an attack coming from $160$K edge-networks is an attack coming
from the entire Internet.

The victim's gateway identifies the $160$K edge-routers that forward attack
traffic and requests that they block their misbehaving clients.
Only Stanford's edge-router $R_1$ cooperates;
$R_2$ to $R_{160K}$ keep forwarding attack traffic.
As a result, the mechanism escalates and the victim's gateway 
uses $159,999$ filters to locally block all traffic coming from 
all edge-networks but Stanford.

When the AITF mechanism completes, the only edge-network
that has preserved its connectivity to ebay.com is Stanford,
and attack traffic has been eliminated.
Of course, by blocking entire edge-networks,
ebay.com also blocks the good traffic coming from them.
However, the alternative is to let attack traffic through,
which would result in most good traffic being dropped
because of congestion.
So, ebay.com benefits from AITF, 
because it preserves a fraction of its good traffic --
even if it is just the good traffic coming from Stanford.
On the other hand, Stanford clearly benefits from AITF, 
because it is the only edge-network to preserve its 
connectivity to the popular website during the attack.

To summarize, there may be millions of attack sources,
but there are only thousands of edge-networks to host them.
A router cannot accommodate millions of filters,
but it does accommodate thousands.
So, the gateway of a DDoS victim may be unable to block
each attack source individually, 
but it \emph{is} able to block each edge-network individually.
However, blocking an entire edge-network results in 
blocking both attack and good traffic coming from it.
To avoid this, the edge-network can take responsibility,
block its misbehaving clients itself,
and preserve its connectivity to the victim.

\subsection{Deployment beyond the Edges}
\label{wide}

We have illustrated that a minimal AITF deployment
at the edges of the Internet
enables the gateway of a DDoS victim to block attack traffic
within a single escalation round.
In this section, we discuss large-scale attacks 
that require wider deployment.

\begin{figure}[htbp]
\centering
\includegraphics[width=8cm, height=4cm]{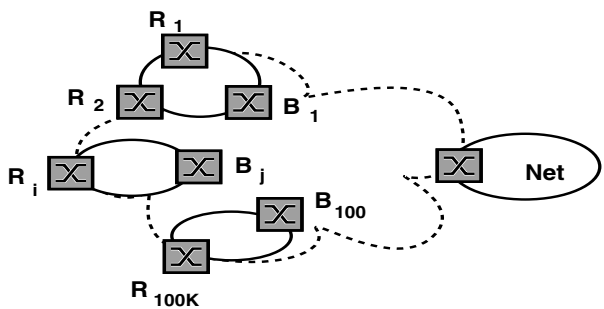}
\caption{\small Wider Deployment - AITF deployed by all border routers in the picture.}
\label{widefig}
\end{figure}

In figure \ref{widefig},
AITF has been deployed throughout the Internet.
A worm compromises $1$M hosts, distributed among $100$K edge-networks,
and commands them to launch DDoS attacks against $100$ $Net$ clients.
At the same time, the worm compromises the attacker gateways,
so that they do not respond to AITF messages.

The victim gateway contacts edge routers $R_1$ to $R_{100K}$ and
requests from each one to block its misbehaving clients.
None of them responds, prompting an escalation.
In the second round, the victim gateway
contacts border routers $B_1$ to $B_{100}$
and requests each one to block all traffic from
its attacker gateways to the victims.
Each border router agrees to cooperate and
sends appropriate filtering requests to its attacker gateways.
The attacker gateways either stop forwarding traffic
to the victims or get disconnected.

In this scenario, it is impossible for $Net$'s gateway
to locally block all traffic from each attacker gateway to each victim --
it would need $100\cdot 100,000 = 10$M filters.
Therefore, $Net$'s gateway requests help from border routers
located close to the attack sources.
This requires (i) that these border routers have deployed AITF, and
(ii) that $Net$'s gateway can inspect a packet and identify,
with high probability, and at wire-speed, the border routers
that forwarded the packet -- which is more than our minimal requirement.

To summarize, our minimal anti-spoofing requirement and 
a minimal AITF deployment at the edges of the Internet
enable a network gateway to block a large-scale attack 
against a few victims.
However, to block multiple large-scale attacks
that potentially involve compromised routers,
a network gateway needs
(i) wider deployment range that includes more than one
border router along the attack path, and
(ii) more information on the attack path, i.e.,
the sequence of border routers that compose it.

The next generation Internet architecture presented in \cite{Cheriton00},
provides such complete path information.
Also, the \emph{Pi} mechanism provides each
packet with an identifier that depends on the \emph{entire}
path followed by the packet~\cite{Yaar03}.
Therefore, we believe that it could be adequately modified
to provide the required path information.
Finally, even if there is no existing solution that provides
the required path information,
we believe that AITF illustrates how useful this information
can be in fighting next-generation DDoS attacks,
and can motivate further research in this direction.

\section{Simulation Results}
\label{sim}

\begin{figure*}[ht]
\centering
\includegraphics[width=5.7cm, angle=-90]{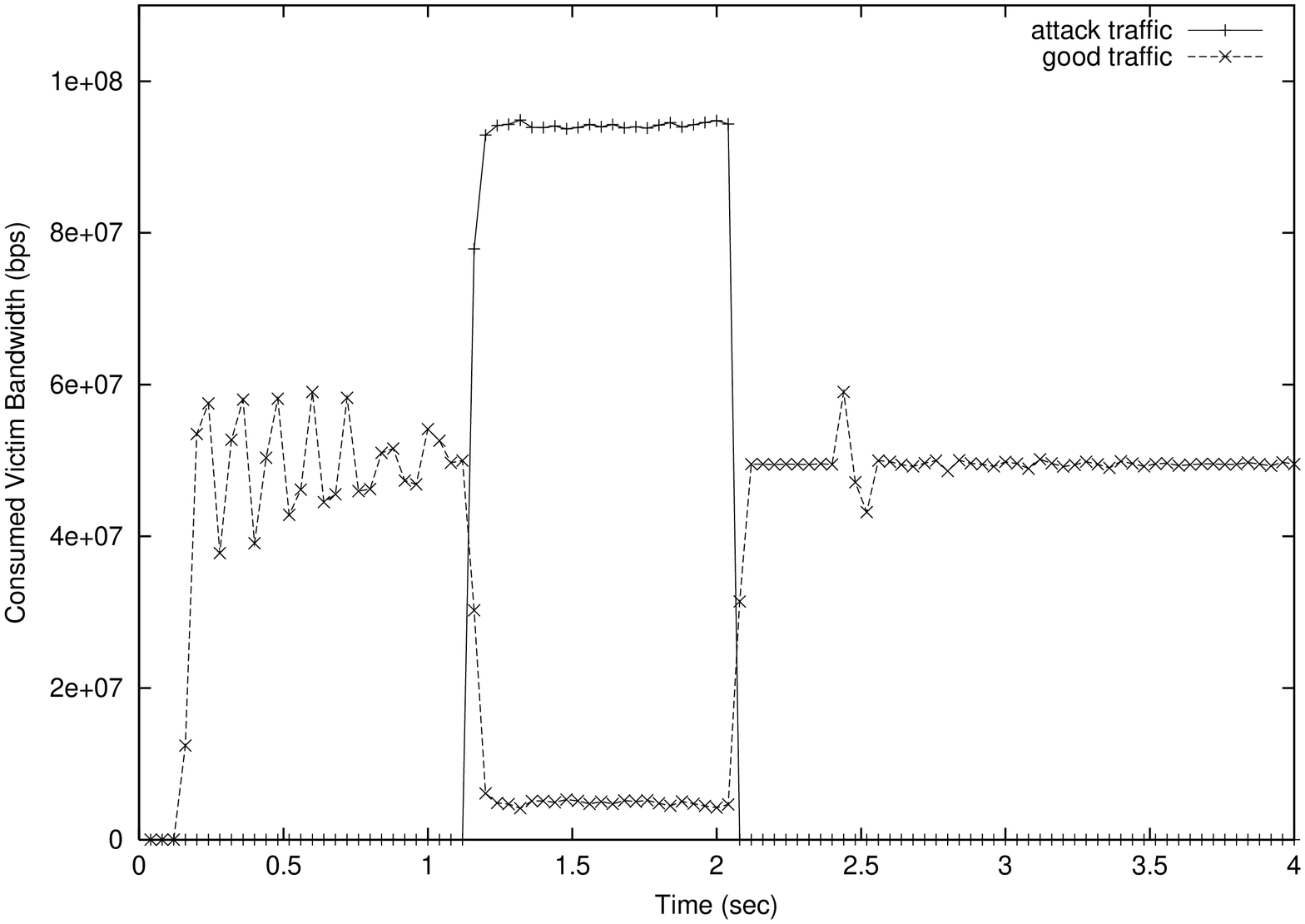}
\includegraphics[width=5.7cm, angle=-90]{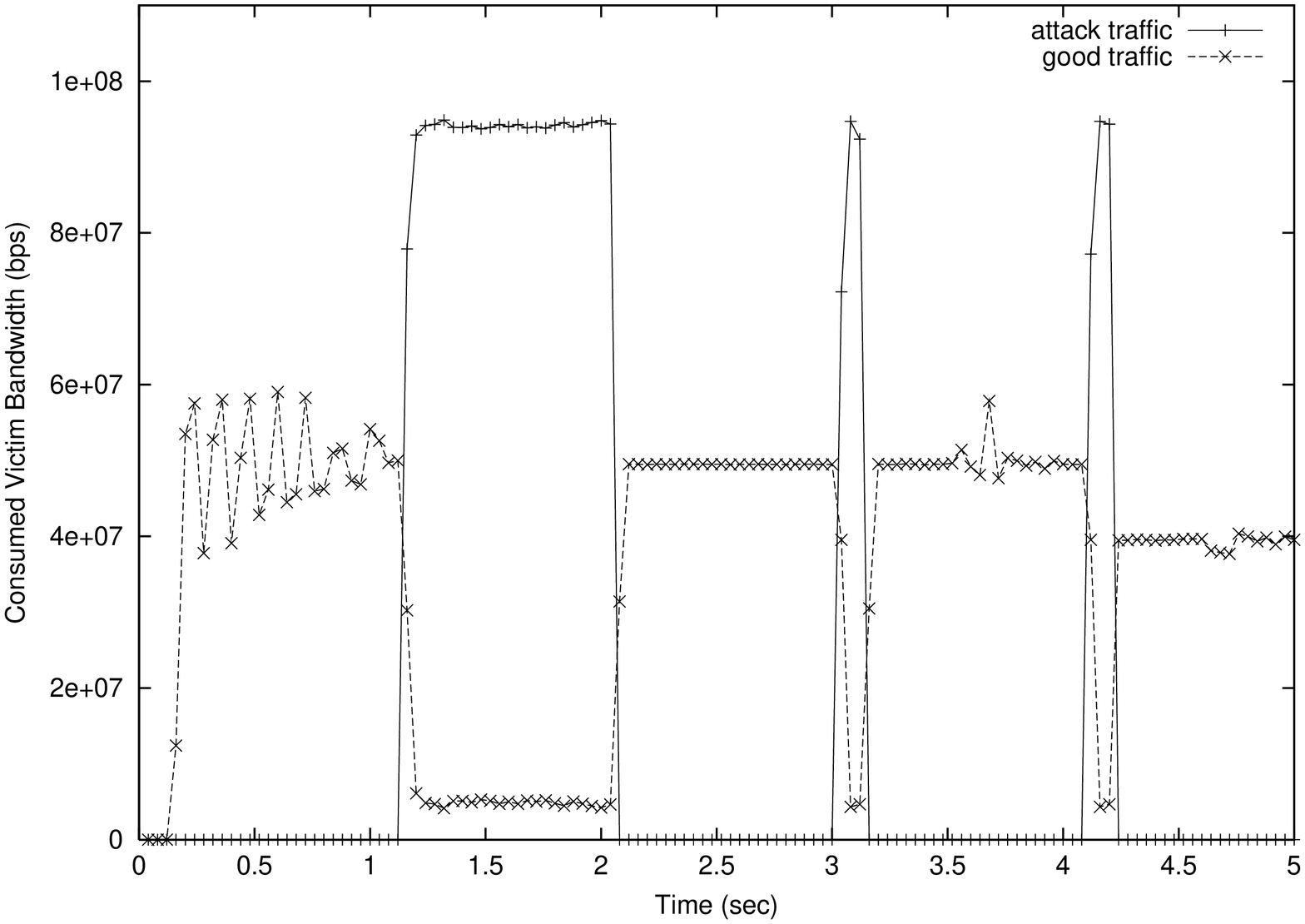}
\caption{\small Mitigating $10$ Gbps attack from $10$K attackers (left);
mitigating same attack through compromised gateways (right).}
\label{simple}
\end{figure*}

\begin{figure*}[ht]
\centering
\includegraphics[width=5.7cm, angle=-90]{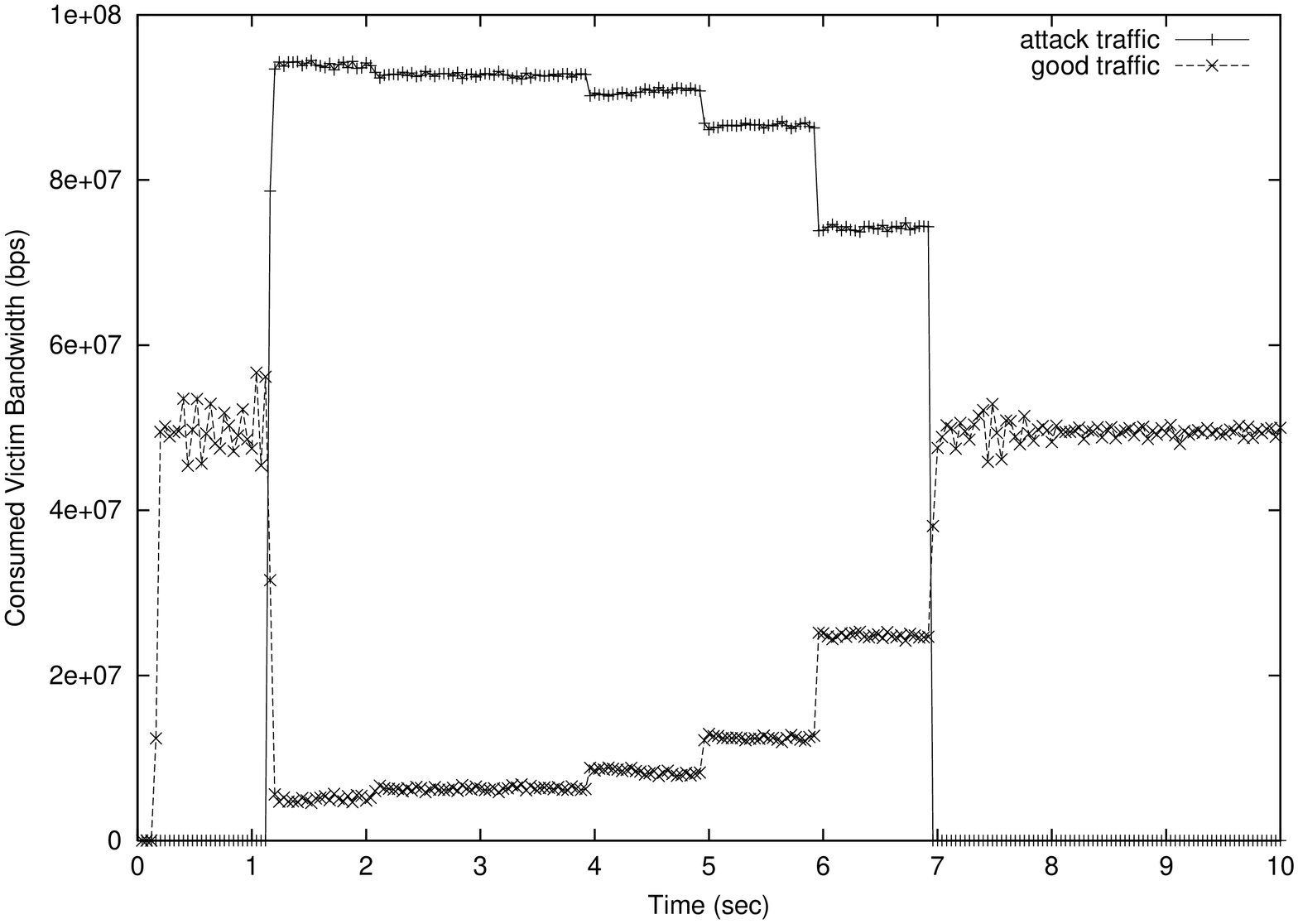}
\includegraphics[width=5.7cm, angle=-90]{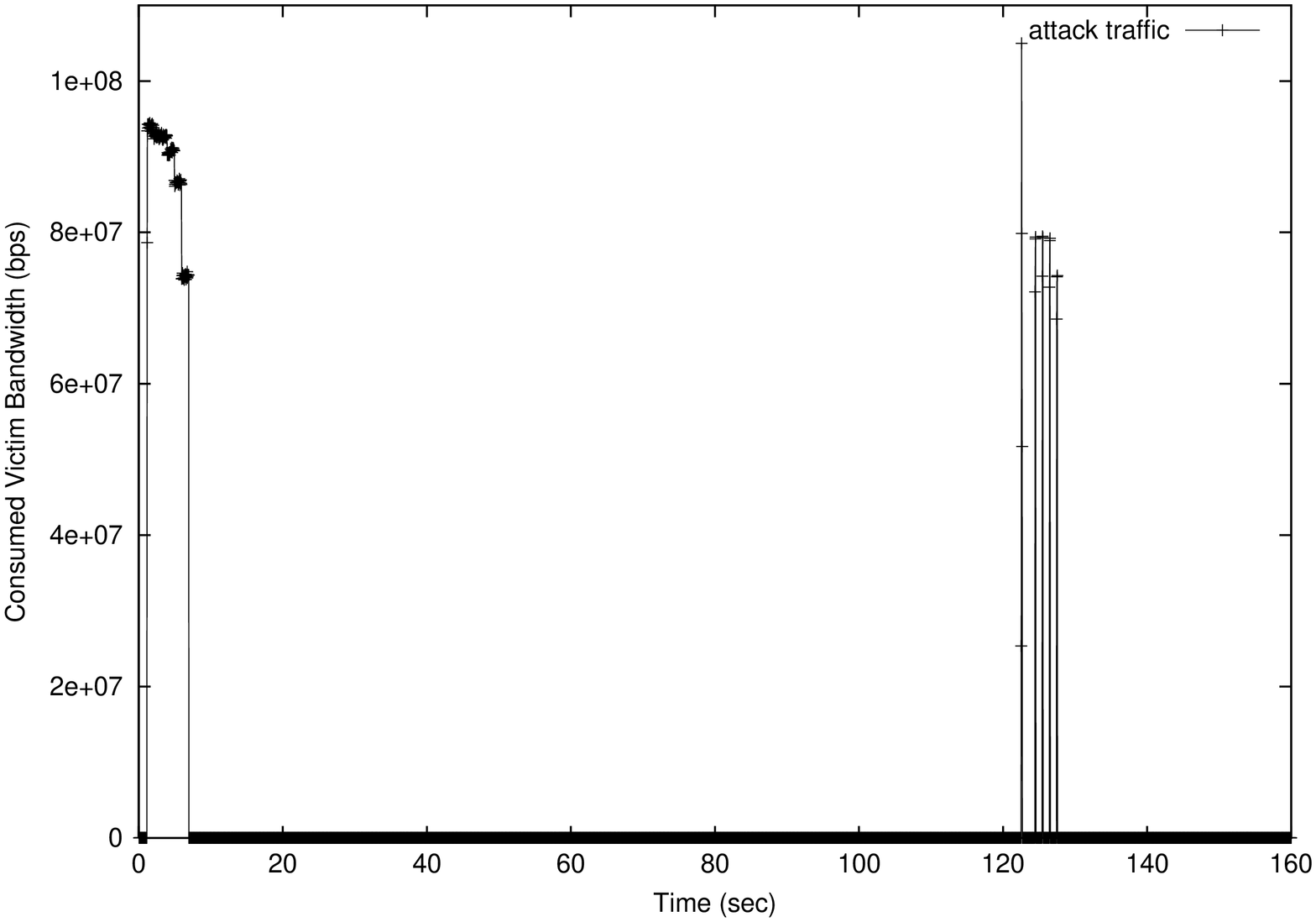}
\caption{\small Mitigating $10$ Gbps attack from $50$K attackers using $10$K filters.
The picture on the left is a zoom in the first $10$ sec.}
\label{light}
\end{figure*}

We built our simulator within the 
Dartmouth Scalable Simulation Framework (DaSSF)~\cite{ssf}.
To create our topology, 
we downloaded Internet routing table data
from the Route Views project site~\cite{routeviews}.
We processed this data using Gao's algorithm
for inferring Autonomous System (AS) relationships~\cite{Gao00}. 

In our simulator, 
we map each AS and each edge-network to a separate AITF domain.
We derive AS topology and peering relationships by applying
Gao's algorithm to the Route Views data.
We derive edge-network topology by roughly creating
one edge-network per advertised class A and class B prefix.\footnote
{We use a simple heuristic to collapse multiple
smaller prefixes into the same edge-network.}
We obtained our routing table data on February 11, 2004;
it yielded $64,608$ AITF domains.\footnote
{$16,825$ ASs, $143,779$ network prefixes, and $56,650$ edge-networks}

Each AITF domain is represented by one AITF router.\footnote{
This leads to worst-case simulation scenarios,
because all AITF messages sent to a domain
are handled by a single router with its limited filtering 
resources.}
AITF routers are interconnected through OC192 ($10$ Gbps) or OC48 ($2.488$ Gbps) links.
End-hosts are connected to their routers through Fast ($100$ Mbps)
or Thin ($10$ Mbps) Ethernet links.
Internet round-trip times average $200$ msec.
Host-to-router round-trip times average $20$ msec.

All our simulation scenarios have the following characteristics:
\begin{itemize}
\item{There is $1$ AITF router acting as the victim's gateway to $10$ DDoS victims.}
\item{Before the attack, each victim is receiving $50$ Mbps of goodput.}
\item{Each victim takes $1$ sec to react to the attack, 
and $100$ msec to detect a recurring flow, 
already identified and blocked in the past.\footnote{
These times are arbitrarily chosen and do not depend on nor affect AITF operation.}}
\item{Each victim is allowed to send up to $1$K filtering requests/sec to its gateway.}
\item{Temporary filter timeout is $T_{tmp} = 1$ sec and the filtering window is $T = 2$ min.}
\item{Attack bandwidths range from $100$ Mbps to $1$ Gbps per victim.}
\item{The number of attackers ranges from $1$K to $300$K per victim,
i.e., $10$K to $3$M overall.}
\end{itemize}

\bigskip

\noindent{\bf Scenario $1$}:
Each victim is sent $1$ Gbps of attack traffic, evenly generated by $1$K attackers.
The attackers are evenly distributed behind $1$K attacker gateways.
All attacker gateways cooperate.

\noindent{\bf Result}:
As shown in figure \ref{simple} on the left,
each victim's goodput is restored within $10$ msec.
The victim's gateway uses overall $10$K filters for $1$ sec.

\noindent{\bf Graph description}:
Before the attack, the victim's goodput is $50$ Mbps;
the attack starts at $t=1$ sec and drives the victim's goodput to $0$;
the victim reacts at $t=2$ sec by sending $1$K filtering requests to its gateway;
the victim's gateway responds within $10$ msec by blocking the $1$K undesired flows.

\bigskip

\noindent{\bf Scenario $2$}:
Each victim is sent $1$ Gbps of attack traffic, evenly generated by $1$K attackers.
The attackers are evenly distributed behind $1$K attacker gateways.
All attacker gateways are compromised and are playing the on-off game
described in \ref{shadow}.

\noindent{\bf Result}:
As shown in figure \ref{simple} on the right,
most of each victim's goodput is restored within $10$ msec, 
with the exception of two $100$ msec spikes.
Some goodput is lost because of escalation.
The victim's gateway uses overall $10$K filters for $2$ min.

\noindent{\bf Graph description}:
The victim's goodput is driven to $0$ when the attack
starts, and restored at $t=2.01$ when the victim's gateway blocks the undesired flows.
Then, the on-off game starts:
The attacker gateways pause forwarding attack traffic at $t = 2.3$, 
when the victim's gateway completes the 3-way handshakes; 
they resume forwarding attack traffic at $t=3.01$,
when the victim's gateway removes the temporary filters.
The victim receives the second attack wave at $t=3.11$,
and reacts at $t=3.21$, by sending a second wave of filtering requests.
This sequence of events is ``replayed'' once more --
recall that each attacker gateway is given two chances --
before the victim's gateway determines that the attacker gateways
are compromised and blocks their traffic locally.
Note that the victim's goodput is not restored $100$\%,
because some of the good traffic gets lost when
the mechanism escalates.

\bigskip

\noindent{\bf Scenario $3$}:
Each victim is sent $1$ Gbps of attack traffic, evenly generated by $5$K attackers.
All attacker gateways cooperate and block attack traffic.
The attackers pause sending undesired traffic when so requested
to avoid disconnection; they resume as soon as the filtering window expires.
Note that the number of undesired flows per victim ($N_{att}=5$K) is higher than the 
maximum per victim filtering request rate ($R=1$K).

\noindent{\bf Result}:
As shown in figure \ref{light},
each victim's goodput is restored within $5$ sec.
The victim's gateway uses $10$K filters for $5$ sec every $2$ min.

\noindent{\bf Graph description}:
The victim reacts at $t=2$ sec
by sending $1$K filtering requests/sec to its gateway.
The victim's goodput is completely restored at $t=7$ sec,
when all $5$K undesired flows are pushed to their gateways.

\bigskip

\noindent{\bf Scenario $4$}:
Each victim is sent $100$ Mbps of attack traffic, evenly generated by $300$K attackers.
All attacker gateways cooperate and block attack traffic.
The attackers pause sending undesired traffic when so requested,
in order to avoid disconnection; they resume as soon as the filtering window expires.
Note that, for each victim, the number of undesired flows ($N_{att}=300$K) is higher
than the number of flows that can be blocked within the filtering window
($R\cdot T = 120$K).

\begin{figure}[htbp]
\centering
\includegraphics[width=5.8cm, angle=-90]{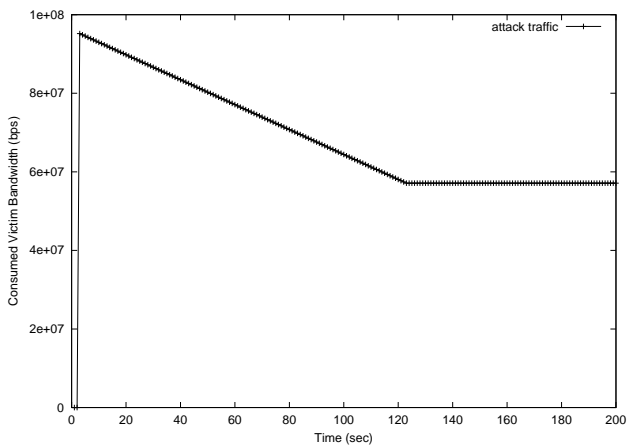}
\caption{\small Mitigating $1$ Gbps attack from $3$M attackers using $10$K filters.}
\label{heavy}
\end{figure}

\noindent{\bf Result}:
As shown in figure \ref{heavy},
$40$\% of each victim's bandwidth is restored within $2$ min.
The victim's gateway uses $10$K filters at all time.

\noindent{\bf Graph description}:
The victim reacts at $t=2$ sec,
by sending $1$K filtering requests/sec to its gateway;
the attackers start resuming their undesired flows at $T=2$ min,
when requests start expiring;
from that point on, for each $1$K undesired flows blocked,
another $1$K are released;
hence, the lost victim bandwidth remains constant.
 
\section{Discussion and Future Work}
\label{issues}

\subsection{Automatic Disconnection is Dangerous}

AITF involves a rather Draconian measure against attack sources:
Either they stop sending undesired traffic or they get disconnected.
This can be abused by malicious node $M$ to disconnect legacy host $L$
that does not understand AITF messages:
\begin{enumerate}
\item{$M$ sends out a filtering request to block all traffic from $L$ to $M$.}
\item{The request reaches $L$'s gateway, which propagates it to $L$
and installs an appropriate filter to verify that $L$ obeys.}
\item{$M$ tricks $L$ into sending it some traffic --
e.g., if $L$ is a web server, $M$ makes an http request to it.}
\item{$L$'s gateway catches the traffic from $L$ to $M$
and disconnects $L$.}
\end{enumerate}
To avoid this abuse, any administrative domain that deploys AITF must
either (i) force its clients to deploy AITF as well,
or (ii) accept the burden of filtering the undesired traffic they forward --
e.g., in the last example, $L$'s gateway does not disconnect $L$,
but installs a filter that blocks all traffic from $L$ to $M$
for $T$ time units.

The second option is less aggressive and more incrementally deployable,
but it disables use of temporary filters and the shadow filtering table
in the attacker's gateway algorithm.
As a result, it requires more filtering resources from each participating gateway,
namely, as many filters as undesired flows generated by the gateway's clients.
As a compromise, a service provider could charge
legacy clients that do not support AITF,
for the potential filtering load induced by their
inability to block their undesired traffic themselves.

\subsection{Compromised Routers}
\label{handabuse}

The 3-way handshake enables a compromised router $cGw$,
which is located on the path between routers $vGw$ and $aGw$,
to interrupt their communication:
\begin{enumerate}
\item{$cGw$ spoofs the IP address of $vGw$ and uses it to send to $aGw$
a SYN packet with a filtering request to block all its traffic to $vGw$.}
\item{$aGw$ responds with a SYN/ACK.}
\item{$cGw$ intercepts the SYN/ACK and completes the handshake.}
\item{$aGw$ blocks all its traffic to $vGw$.}
\end{enumerate}

On the other hand,
a compromised router on the path between two victim nodes
can disrupt the victims' communication anyway, with or without AITF,
e.g. by dropping all their traffic.
Of course, AITF does offer a compromised router yet another way to
damage transit traffic.
Nevertheless, the 3-way handshake is necessary for scalability,
because it enables edge-routers to directly establish filtering
agreements, bypassing the Internet core.

\subsection{Forged Requests}
\label{forged}

As described so far, AITF can be abused 
by malicious node $M$ to interrupt the communications
of node $N$ that is located on the same subnet with $M$, 
i.e., to harm the connectivity \emph{of its own} subnet:
\begin{enumerate}
\item{$M$ spoofs the IP address of $N$ and sends a filtering
request to block certain traffic addressed to $N$.}
\item{$M$'s gateway cannot verify that the filtering request is spoofed
and completes the appropriate 3-way handshakes.}
\item{$N$ looses part (or all) of its communications.}
\end{enumerate}
To avoid this abuse, any administrative domain that deploys AITF
must either (i) prevent source address spoofing \emph{in its own network}
or (ii) authenticate victim filtering requests -- 
i.e., requests coming from its own clients.

\subsection{DDoS against AITF}
\label{selfddos}

The only part of the AITF mechanism that is susceptible to
a DDoS attack is the attacker's gateway algorithm,
because it is the only one to accept requests from unknown sources.
A router running the victim's gateway algorithm
only accepts rate-limited requests from its own customers.
Similarly, an end-host/router suspected of being an attacker,
only accepts rate-limited requests from its own providers --
a correctly functioning provider would not overload a customer with filtering requests
and then disconnect the customer for failing to satisfy them.

The attacker's gateway algorithm is susceptible to the
following attack:
If AITF router $Gw$ is flooded with AITF messages
sent by alleged victim gateways,
it spends all its resources trying to process them
and fails to satisfy the ``real'' ones.
Note that, in order for this to be a problem,
there must be ``real'' AITF messages,
i.e., there must be $Gw$ clients actually sending undesired flows.
Therefore, in order to launch such an attack on $Gw$,
an attacker must compromise (i) enough Internet hosts to flood $Gw$
with bogus filtering requests and (ii) enough $Gw$ clients
to actually start an attack behind $Gw$.
Such a powerful attacker can indeed affect $Gw$'s
ability to execute the attacker's gateway algorithm.

$Gw$ can take two steps to mitigate such an attack:
(i) Avoid local buffering of SYN messages corresponding
to uncompleted handshakes.
Instead, use the SYN cookie technique \cite{cookie}
employed by TCP to prevent SYN flooding attacks.
(ii) Inspect the received AITF messages
and identify the AITF routers that forward an excess of messages
and filter/rate-limit them
(recall that, according to our minimal anti-spoofing requirement,
a node can inspect a packet and identify, 
with high probability, and at wire-speed,
a border router that forwarded the packet and is located close to its source).
Ultimately, if the misbehaving AITF routers are too many to filter locally,
$Gw$ can act as a DDoS victim and use itself AITF
to push filtering of undesired traffic close to its sources.
However, we have not yet studied nor simulated any scenario where 
an AITF router acts as a victim.

\subsection{Dynamic Allocation of Filtering Resources}
\label{dynamic}

An AITF provider does not have to statically preallocate a fixed number 
of filters to every client.
When there are enough filters, there is no reason to deny satisfying
a filtering request, even when the corresponding client has exceeded
its maximum rate.
In the same way, when the provider is running out of filters,
it can dynamically communicate to the clients to lower their
filtering request rates.

\section{Related Work}
\label{related}

Having briefly introduced the Pushback mechanism in section \ref{intro},
we now make a more detailed comparison to AITF.

Pushback assumes no anti-spoofing mechanism and, thus,
defines an attack flow only in terms of its IP destination prefix.
For example, if end-host $V$ is under attack, 
$V$'s edge-router identifies the link that contributes most to the attack
and blocks (or rate-limits) all traffic arriving on that link and
addressed to $V$.
As a result, Pushback (i) is directly deployable in the current Internet,
and (ii) requires from each router only $1$ filter per victim.
On the other hand,
if attack traffic is uniformly distributed across the inbound links, 
a pushback router ends up sacrificing all (or most) good traffic to the victim --
an effect characterized in \cite{Mahajan01} as \emph{collateral damage}.
As a result, Pushback is not effective when the attack sources
are evenly distributed across the Internet.

Pushback propagates filtering requests
hop-by-hop through the network.
As a result, Pushback does not need to verify the authenticity
of filtering requests.
On the other hand, hop-by-hop propagation of filtering requests
places a large load on the core of the Internet,
turning it into a potential ``filtering bottleneck''.
The current Pushback design avoids this problem
by filtering based on the IP destination prefix --
which requires only $1$ filter per victim.
However, this limits Pushback's ability 
to selectively block attack traffic and eliminate collateral damage,
even in presence of an anti-spoofing mechanism.
We believe that source address spoofing is 
a serious enough problem to motivate the 
deployment of an Internet-wide anti-spoofing mechanism.
AITF was designed to leverage such a mechanism
to avoid collateral damage and block large numbers of
concurrent attack flows as close as possible to their sources.

Another mechanism for mitigating DDoS attacks is the 
Stateless Internet Flow Filter (SIFF)~\cite{Yaar04}.
In this scheme, all Internet traffic is divided into two categories:
Privileged and non-privileged.
Privileged traffic always receives priority over non-privileged traffic.
A client establishes a privileged channel to a server through
a capability exchange handshake that involves packet marking by 
all the routers on the path;
the client includes the capability in each packet it sends to the server;
each router along the path verifies the capability and gives
priority to the client's traffic.
The main advantages of SIFF are that
(i) it does not require any filtering state in the routers,
and (ii) it does not assume any cooperation between ISPs.
On the other hand, it requires deployment in all the routers
on the client-to-server path that control bottleneck links.
Also, once a server is under attack, 
a new client must try multiple times 
(the exact number depends on the strength of the attack)
to establish a privileged channel,
because channel establishment unavoidably involves an exchange 
of two non-privileged packets.
Finally, it is possible for malicious nodes to establish
privileged connections between them and flood the network
with privileged traffic -- though it should be noted
that this attack requires cooperating attacker pairs.

Lakshminarayanan {\it et al.} have proposed using the Internet 
Indirection Infrastructure (I3) to enable
a victim to stop certain types of traffic by removing the corresponding
I3 identifier from the network~\cite{Lakshmi03}.
This mechanism can be used to block traffic addressed to unutilized ports,
contain the traffic of individual applications, 
or prioritize the traffic of already established connections.
It does not address the issue of blocking a large number of 
undesired flows using a reasonable number of filters per router.

\section{Conclusions}
\label{conclusion}

We presented AITF, a mechanism that protects public-access sites
against highly distributed denial-of-service attacks.
Its main advantages are that 
(i) it blocks large numbers of attack flows,
while requiring from each participating router only a reasonable
number of filters, and
(ii) it is incrementally deployable.

More specifically, if AITF is widely deployed,
an AITF router with $10$K filters can
mitigate an attack coming from $3$M sources 
uniformly distributed across the Internet, 
by restoring $40$\% of the victim's bandwidth within $2$ minutes. 
If none of the edge-networks hosting attack sources has deployed AITF,
the victim's gateway still achieves the same result,
but needs up to $160$K filters (as many as all the Internet edge-networks).
If an attack source is located in an edge-network that has not deployed AITF,
blocking its traffic comes at the cost of blocking all traffic from the
specific edge-network to the victim.
Only the edge-networks that have deployed AITF
preserve their connectivity to the victim.

The simple idea behind AITF is that the Internet does have enough
filtering capacity to block large amounts of undesired traffic --
it is just that this capacity is concentrated close to the undesired traffic sources.
AITF enables service providers to ``gain access'' to this filtering
capacity and couple it with a reasonable amount of their own filtering resources,
in order to protect their customers in the face of increasingly
distributed denial-of-service attacks.

\bibliographystyle{plain}
\bibliography{references}

\appendix

\section{Appendix}
\label{proof}

\myitem{Suppose the victim sends filtering requests for $X$ 
undesired flows $F^i$, $i=1..X$.
Filtering response time for $F^i$ is $T_{\mathit {fr}}^i$.}

\myitem{The bandwidth consumed by $F^i$ is \\
$B_{\mathit {l}}^i = B_{\mathit {att}}^i \cdot\frac{T_{\mathit {fr}}^i}{T+T_{\mathit {fr}}^i} 
\approx B_{\mathit {att}}^i \cdot\frac{T_{\mathit {fr}}^i}{T}$.}

\myitem{Aggregate bandwidth consumed by the $X$ undesired flows is 
$B_{\mathit {l}}^x \le B_{\mathit {att}}^x \cdot\frac{T_{\mathit {fr}}}{T}$, 
where $B_{\mathit {att}}^x$ is the original aggregate bandwidth of the $X$ flows
and $T_{\mathit {fr}} = max{T_{\mathit {fr}}^i}$.}

\myitem{Suppose the victim is receiving $N_{\mathit {att}}$ undesired flows,
but sends filtering requests only for $X$ of them.}

\myitem{Aggregate bandwidth consumed by the $N_{\mathit {att}}$ flows is
$B_{\mathit {l}} = B_{\mathit {att}} - B_{\mathit {att}}^x + B_{\mathit {l}}^x \\
= B_{\mathit {att}} - B_{\mathit {att}}^x (1 - \frac{T_{\mathit {fr}}}{T}) 
\approx B_{\mathit {att}} - B_{\mathit {att}}^x \\
\le B_{\mathit {att}}(1-\frac{X}{N_{\mathit {flows}}})$.}

\myitem{The victim is allowed $R$ filtering requests per time unit.
Thus, it can keep a total of $RT$ undesired flows blocked.}

\myitem{Substituting $X$ for $RT$, lost victim bandwidth is
$B_{\mathit {l}} \le B_{\mathit {att}}(1-\frac{RT}{N_{\mathit {att}}})$.}

\end{document}